\documentclass[pre,aps,12pt]{revtex4}
\usepackage{graphicx,color}

\newcommand{\FeGd}{\protect Fe$_{10}$Gd$_{10}$}
\newcommand{\eqref}[1]{Eq.~(\protect\ref{#1})}

\begin{document}

\title{The delta-chain with ferro- and antiferromagnetic
interactions in applied magnetic field}
\author{D.~V.~Dmitriev}
\author{V.~Ya.~Krivnov}
\email{krivnov@deom.chph.ras.ru}
\affiliation{Institute of
Biochemical Physics of RAS, Kosygin str. 4, 119334, Moscow,
Russia.}
\author{J.~Schnack}
\affiliation{Fakult\"at f\"ur Physik, Universit\"at Bielefeld,
Postfach 100131, D-33501 Bielefeld, Germany}
\author{J.~Richter}
\affiliation{Institut f\"{u}r Physik,
Otto-von-Guericke-Universit\"{a}t Magdeburg, P.O. Box 4120, 39016
Magdeburg, Germany} \affiliation{Max-Planck-Institut f\"{u}r
Physik komplexer Systeme, N\"{o}thnitzer Stra\ss e 38, 01187
Dresden, Germany}
\date{}

\begin{abstract}
We study the thermodynamics of the delta-chain with competing
ferro- and antiferromagnetic interactions in an external magnetic
field which generalizes the field-free case studied previously.
This model plays an important role for the recently synthesized
compound Fe$_{10}$Gd$_{10}$ which is nearly quantum critical. The
classical version of the model is solved exactly and explicit
analytical results for the low-temperature thermodynamics are
obtained. The spin-$s$ quantum model is studied using exact
diagonalization and finite-temperature Lanzos techniques.
Particular attention is focused on the magnetization and the
susceptibility. The magnetization of the classical model in the
ferromagnetic part of the phase diagram defines the universal
scaling function which is valid for the quantum model. The
dependence of the susceptibility on the spin quantum number $s$ at
the critical point between the ferro- and ferrimagnetic phases is
studied and the relation to Fe$_{10}$Gd$_{10}$ is discussed.
\end{abstract}

\maketitle

\section{Introduction}
\label{intro}
\begin{figure}[tbp]
\includegraphics[width=5in,angle=0]{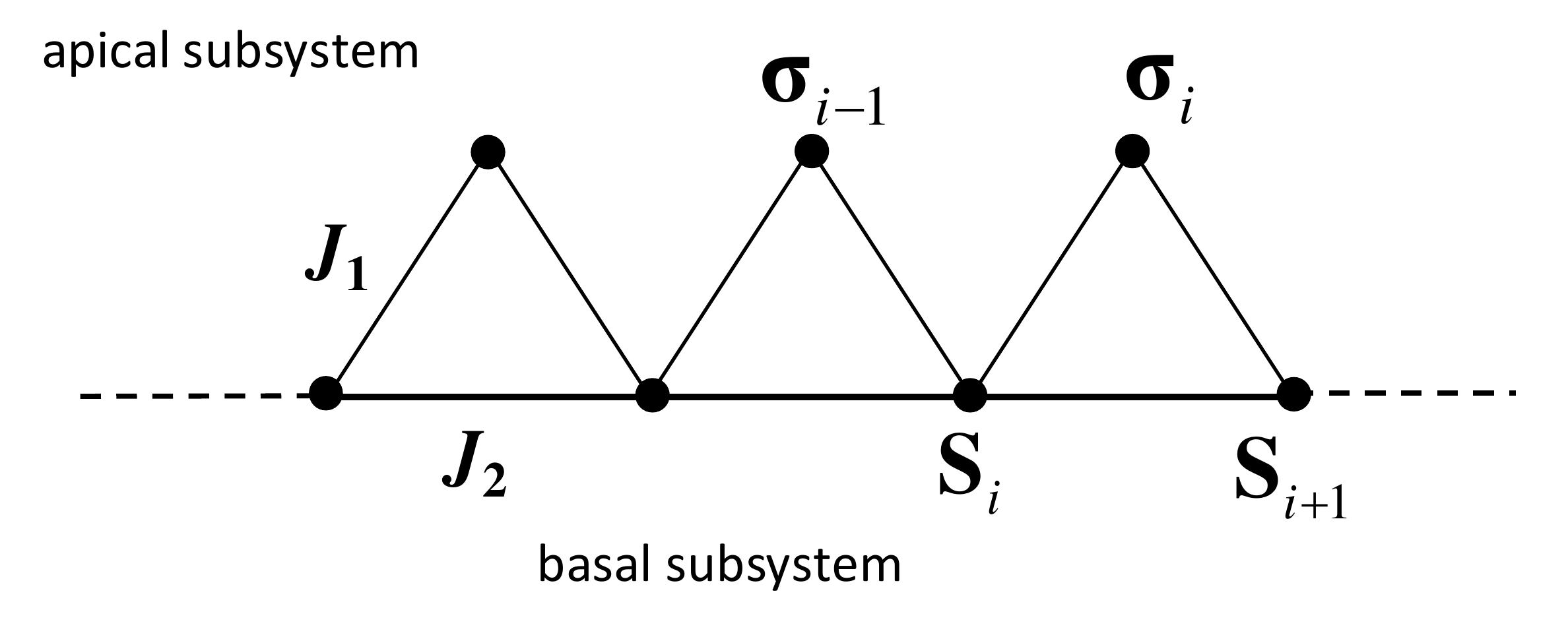}
\caption{The delta-chain model.}
\label{Fig_saw}
\end{figure}

Low-dimensional quantum magnets on geometrically frustrated
lattices have been extensively studied during last years
\cite{diep,mila,qm}. One of the interesting classes of such systems
includes lattices consisting of triangles.
%There is a broad class of highly frustrated spin systems
%having triangular geometry.
A typical example of these objects is the delta or the sawtooth
chain, i.e. a Heisenberg model defined on a linear chain of
triangles as shown in Fig.~\ref{Fig_saw}. The Hamiltonian of
this model has the form:
\begin{equation}
\hat{H}=J_{1}\sum \mathbf{\sigma }_{i}\cdot (\mathbf{S}_{i}+\mathbf{%
S}_{i+1})+J_{2}\sum \mathbf{S}_{i}\cdot \mathbf{S}%
_{i+1}-H\sum (\sigma _{i}^{z}+S_{i}^{z}),  \label{Hq}
\end{equation}%
where $\sigma _{i}$ and $S_{i}$ are the apical and the basal spins
correspondingly, $H$ is the external magnetic field, $J_{1}$ and
$J_{2}$ are apical-basal and basal-basal interactions and a direct
interaction between the apical spins is absent.

The quantum $s=\frac{1}{2}$ delta-chain with antiferromagnetic
(AF) exchange interactions $J_{1}$ and $J_{2}$ ($J_{1},J_{2}>0$)
has been studied extensively and it exhibits a variety of peculiar
properties \cite{sen,nakamura,blundell,flat,Zhit,prl,Derzhko2004}.
At the same time the $s=\frac{1}{2} $ delta-chain with
ferromagnetic $J_{1}$ and antiferromagnetic $J_{2}$ interaction
(F-AF delta-chain) is very interesting as well and has unusual
properties depending on the frustration parameter $\alpha
=\frac{J_{2}}{\left\vert J_{1}\right\vert }$
\cite{Tonegawa,Kaburagi,KDNDR,DK15}. In particular, the ground
state of this model is ferromagnetic for $\alpha <\frac{1}{2}$ and
it is believed \cite{Tonegawa} that it is ferrimagnetic for
$\alpha>\frac{1}{2}$. The critical point $\alpha =\frac{1}{2}$ is
the transition point between these two ground state phases. The
ground state properties of the model in this point are highly
non-trivial. For example, the $s=\frac{1}{2}$ F-AF delta-chain
studied in Ref.~\cite{KDNDR} has a class of localized magnon bound
states which form a macroscopically degenerate ground state
manifold hosting already half of the maximum total entropy $N\ln
2$. The $s=\frac{1}{2}$ F-AF delta-chain is a minimal model for a
description of real compounds, in particular malonate-bridged
copper complexes \cite{Inagaki,Tonegawa,ruiz,Kaburagi} as well as
the new kagome fluoride Cs$_2$LiTi$_3$F$_{12}$, that hosts F-AF
delta-chains as magnetic subsystems \cite{SUJ:PRB19}.

The $s=\frac{1}{2}$ F-AF model can be extended to the delta-chain
composed of two types of spins
($\mathbf{\sigma}_{i},\mathbf{S}_{i}$) characterized by the spin
quantum numbers $S_{a}$ and $S_{b}$ of the apical and basal spins,
respectively. The ground state of this model is ferromagnetic (F)
for $\alpha<\alpha_{c}$ and non-collinear ferrimagnetic for
$\alpha>\alpha_{c}$, where $\alpha_{c}=S_{a}/2S_{b}$. The ground
state of the model with any quantum numbers $S_{a}$ and $S_{b}$ in
the critical point $\alpha_{c}$ consists of exact multi-magnon
states as for the $s=\frac{1}{2}$ model and has similar
macroscopic degeneracy \cite{KDNDR}.

An additional motivation for the study of the ($S_{a},S_{b}$) F-AF
delta-chain is the existence of a recently synthesized mixed $3d/4f$
cyclic coordination cluster
[Fe$_{10}$Gd$_{10}$(Me-tea)$_{10}$(Me-teaH)$_{10}$(NO$_{3}$)$_{10}$]20MeCN
(i.e. \FeGd) \cite{S60}. This cluster consists of
$10+10$ alternating gadolinium and iron ions and its spin
arrangement corresponds to the delta-chain with Gd and Fe ions
as the apical and basal spins correspondingly. As it was
established in Ref.~\cite{S60} that the exchange interaction
between neighboring Fe ions is antiferromagnetic
($J_{2}\simeq 1.3K$) and the
interaction between neighboring Fe and Gd is ferromagnetic
($J_{1}\simeq -2.0K$). The spin values of Fe and Gd ions are $S=\frac{5}{2}$
for Fe$^{\text{III}}$ and $S=\frac{7}{2}$ for Gd$^{\text{III}}$, respectively. The ground
state spin of this cluster is $S=60$ which is one of the largest
spins of a single molecule \cite{Sch:CP19}.
This molecule is a finite-size
realization of the F-AF delta-chain with $S_{a}=\frac{7}{2}$ and
$S_{b}=\frac{5}{2}$. Remarkably, according to the estimate of the
values of $J_{1}$ and $J_{2}$ in Ref.~\cite{S60} the frustration
parameter is $\alpha=0.65$, i.e. it is very close to the critical
value of $\alpha_{c}=0.7$. Therefore, this molecule, although it
is not directly at the critical point and located in the F phase,
has properties which are strongly influenced by the nearby quantum
critical point. Because the spin quantum numbers for Fe and Gd
ions are rather large it seems that the classical approximation
for the ($S_{a},S_{b}$) F-AF delta-chain is justified.
%Therefore, the
%study of the classical F-AF delta-chain is relevant.

In our preceding work \cite{DKRS} we study the classical version
of the F-AF delta-chain at zero magnetic field. The ground state
phase diagram of the classical model consists of the ferromagnetic
at $\alpha <\alpha_{c}$ and the ferrimagnetic at $\alpha
>\alpha_{c}$ phases. Remarkably, the transition between
these phases occurs at the same frustration parameter $\alpha_c$
as in the quantum model. For $S_{a}=S_{b}$ the critical point
between the ferromagnetic and ferrimagnetic phases is at
$\alpha=\frac{1}{2}$. In Ref.~\cite{DKRS} we have obtained exact results
for the partition function, the thermodynamics and spin
correlation functions for different regions of the parameter
$\alpha $. It was shown that the classical model provides a
reasonable description of thermodynamics of \FeGd\ down to
moderate temperatures. In Ref.~\cite{DKRS} we have
studied also quantum corrections to the classical results which
are essential at low temperature. It was shown that some
properties of the quantum spin delta-chain are correctly described
by the classical model. For example, the main features of the
zero-field susceptibility $\chi$ of the quantum spin delta-chain
are reproduced by the classical model. In particular, the behavior
of the susceptibility in the F phase (at $\alpha <\alpha_{c}$) of
the classical model coincides with the quantum model in both low
and high temperature limits. The product $\chi T$ per spin
diverges as $T^{-1}$ at $T\to 0$ in the infinite chain and it is
proportional to $N$ for finite system and such a dependence of $\chi
T$ takes place, in particular, in \FeGd.
However, the results of the paper \cite{DKRS} are related to the
zero field case. The experimental data for \FeGd\
presented in Ref.~\cite{S60} demonstrate that there is a strong
influence of a magnetic field on the low-temperature
thermodynamics. That is related to the massively degenerate
manifold of localized magnon states having different total
magnetization. The Zeeman term will partly lift this degeneracy,
this way influencing the low-energy spectrum substantially.
Therefore, it is interesting to consider the thermodynamic
behavior of the classical delta-chain in a magnetic field. In
this paper we will study the classical delta-chain in the external
magnetic field. This model is more complicated in comparison with
that for $H=0$. Nevertheless, it can be solved exactly and the
analytical results for the low-temperature properties are obtained
explicitly. We calculate the magnetization curve $M(H)$ and the
susceptibility and compare them with the results for the quantum
model. For example, we can quantitatively explain the experimental
result related to a maximum of $MT/H$ vs. $T$ for \FeGd.

For simplicity and to avoid cumbersome formulas we will consider
the spin-$s$ delta-chain, i.e. the model with $S_{a}=S_{b}=s$.
(The extension of results for the case $S_{a}\neq S_{b}$ can be
obtained straightforwardly). In accordance with the adopted
simplification we will further consider the F-AF delta-chain with
$s=3$ as a model for the \FeGd\ molecule.

The paper is organized as follows. In Sec.~IIA we describe the
ground state of the classical model (\ref{H}) in different regions
of the frustration parameter $\alpha $ including the critical
value $\alpha =\frac{1}{2}$. The partition function and the
magnetization are calculated in Sec.~IIB. In Sec.~IIC explicit
analytical results in the low-temperature limit are presented for
different regions of the parameter $\alpha $ and the scaling law
for $\alpha \leq \frac{1}{2}$ is established. In Sec.~III the
quantum effects at low temperatures will be studied by a
combination of full exact diagonalization (ED) using J.
Schulenburg's \textit{spinpack} code \cite{spinpack} and the
finite temperature Lanczos (FTL) technique \cite{FTL1,FTL2} . We
compare the magnetization of the classical and the quantum models
and estimate finite-size effects.

\section{Classical spin $\Delta $-chain in a magnetic field}

To obtain the classical version of Hamiltonian (\ref{Hq}) we set
$\sigma _{i}=s\vec{n}_{i}$ and $S_{i}=s\vec{n}_{i}$, where
$\vec{n}_{i}$ is the unit vector at the $i$-th site. Taking the
limit of infinite $s$ we arrive at the Hamiltonian of the
classical delta-chain
\begin{equation}
\mathcal{H}=-\sum_{i=1}^{N}\vec{n}_{i}\cdot \vec{n}_{i+1}+\alpha \sum_{i=1}^{N/2}\vec{n}%
_{2i-1}\cdot \vec{n}_{2i+1}-h\sum_{i=1}^{N}n_{i}^{z},  \label{H}
\end{equation}%
where $N$ is the number of spins. In \eqref{H} we take the
apical-basal interaction as $-1$ and the basal-basal interaction
as $\alpha $.

In this Section we use the normalized magnetic field and
temperature
\begin{eqnarray}
h&=&H/s \\
t&=&T/s^2  \label{beta}
\end{eqnarray}%
and the corresponding inverse temperature $\beta =1/t$ to present
the thermodynamic properties of model (\ref{H}).

\subsection{Ground state}

We start our study of model (\ref{H}) from the determination of
the ground state. For this aim it is useful to represent
Hamiltonian (\ref{H}) as a sum over triangle Hamiltonians
\begin{equation}
\mathcal{H}=\sum_{i=1}^{N/2}\mathcal{H}_{\Delta }(i)  \label{Hsum}
\end{equation}%
where the Hamiltonian of $i$-th triangle has the form
\begin{equation}
\mathcal{H}_{\Delta }(i)=-\vec{n}_{2i-1}\cdot \vec{n}_{2i}-\vec{n}_{2i}\cdot \vec{n}%
_{2i+1}+\alpha \vec{n}_{2i-1}\cdot \vec{n}_{2i+1}-\vec{h}\cdot (\frac{1}{2}%
\vec{n}_{2i-1}+\vec{n}_{2i}+\frac{1}{2}\vec{n}_{2i+1}).
\label{Htri}
\end{equation}
To determine the ground state of model (\ref{Hsum}) we need to
find the spin configuration on each triangle which minimizes the
classical energy. It turns out that the lowest spin configuration
on a triangle is different in the regions $\alpha \leq \frac{1}{2}$
and $\alpha >\frac{1}{2}$. For $\alpha \leq \frac{1}{2}$ the
ground state is the trivial ferromagnetic one with all spins on
each triangle pointing in the same direction. The global spin
configuration of the whole system in this case is obviously
ferromagnetic as well.

\begin{figure}[tbp]
\includegraphics[width=5in,angle=0]{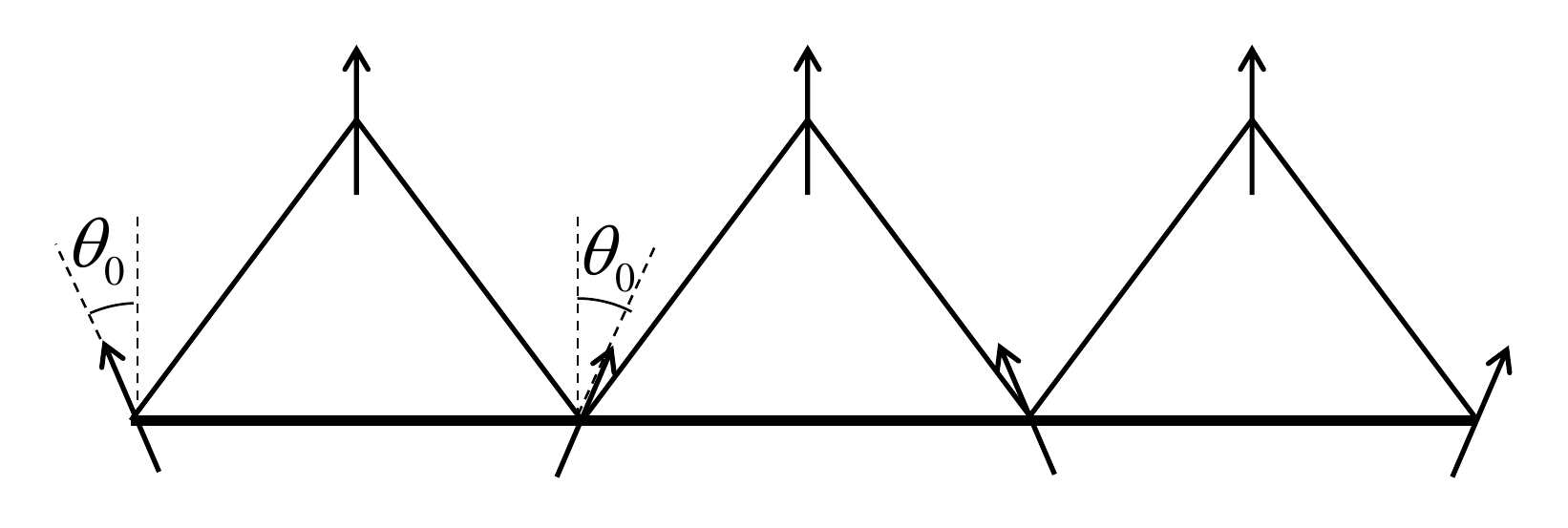}
\caption{The ferrimagnetic ground state of classical delta-chain.}
\label{Fig_ferri}
\end{figure}

For $\alpha >\frac{1}{2}$ the lowest classical energy on each
triangle is given by a non-collinear ferrimagnetic configuration,
where all spins of triangle $\vec{n}_{1},\vec{n}_{2},\vec{n}_{3}$
lie in the same plane and spin $\vec{n}_{2}$ assumes an equal angle
$\theta _{0}$ with spins $\vec{n}_{1}$ and $\vec{n}_{3}$. The
global ground state without magnetic field of the whole system for
$\alpha >\frac{1}{2}$ is macroscopically degenerate
\cite{Chandra}. The magnetic field lifts this degeneracy and
stabilizes the ferrimagnetic configuration where all apical spins
are directed along the magnetic field and the basal spins are
inclined by an equal angle $\theta _{0}$ to the right and to the
left of the field direction as shown in Fig.~\ref{Fig_ferri}. Therefore,%
\begin{eqnarray}
\vec{h}\cdot \vec{n}_{2} &=&h \\
\vec{n}_{1}\cdot \vec{n}_{2} &=&\vec{n}_{2}\cdot \vec{n}_{3}=\cos \theta _{0}
\\
\vec{n}_{1}\cdot \vec{n}_{3} &=&\cos \left( 2\theta _{0}\right) \\
\cos \theta _{0} &=&\frac{2+h}{4\alpha }.  \label{theta0}
\end{eqnarray}
The magnetization of the ground state in the ferrimagnetic
region is
\begin{equation}
M_{\mathrm{gs}}=\frac{2\alpha +1}{4\alpha }+\frac{h}{8\alpha }  \label{Mgs}
\end{equation}%
for $h<h_{\mathrm{sat}}$, where the saturated field in the ground
state is defined by condition $\theta_0=0$:
\begin{equation}
h_{\mathrm{sat}}=4\alpha-2. \label{hsat}
\end{equation}

\subsection{Partition function}

The partition function $Z$ of model (\ref{H}) is%
\begin{equation}
Z=(\prod_{i=1}^{N}\int d\vec{n}_{i})\exp \left( -\beta \mathcal{H}
\right). \label{Z}
\end{equation}
In our previous paper \cite{DKRS} we used local coordinate systems
associated with the $i$-th spin, which substantially simplified
calculations. For the system in a magnetic field this trick does not
work. Therefore, we follow a common transfer-matrix method which
reduces the calculation of the partition function in 1D systems to
an integral equation \cite{Blume,Harada}. In our case this
integral equation is written for one triangle and has the form:
\begin{equation}
\int e^{-\beta \mathcal{H}_{\Delta }(1)}\psi _{i}(\vec{n}_{1})d\vec{n}_{2}d\vec{n}%
_{1}=\lambda _{i}\psi _{i}(\vec{n}_{3}).  \label{IntEq}
\end{equation}
The eigenvalues $\lambda _{i}$ define the partition function as%
\begin{equation}
Z=\sum \lambda _{i}^{N/2}.
\end{equation}
In the thermodynamic limit $N\to\infty $ only the largest eigenvalue $%
\lambda_0$ survives:
\begin{equation}
Z\to \lambda _{0}^{N/2}.
\end{equation}
Selecting the terms containing the apical spin $\vec{n}_{2}$ in the
Hamiltonian of one triangle (\ref{Htri})%
\begin{equation}
\mathcal{H}_{\Delta }(1)=-\vec{n}_{2}\cdot \left( \vec{n}_{1}+\vec{n}_{3}+\vec{h}%
\right) +\alpha \vec{n}_{1}\cdot
\vec{n}_{3}-\frac{1}{2}\vec{h}\cdot \left(
\vec{n}_{1}+\vec{n}_{3}\right)
\end{equation}%
we can explicitly integrate the integral equation (\ref{IntEq}) over
the apical spin $\vec{n}_{2}$
\begin{equation}
\int d\vec{n}_{2}\exp \left[ \beta \vec{n}_{2}\cdot \left( \vec{n}_{1}+\vec{n%
}_{3}+\vec{h}\right) \right] =\frac{\sinh \left( \beta
h_{a}\right) }{\beta h_{a}},
\end{equation}%
where $h_{a}$ is the effective magnetic field acting on the apical
spin $\vec{n}_{2}$:
\begin{equation}
h_{a}=\sqrt{\left( \vec{n}_{1}+\vec{n}_{3}+\vec{h}\right) ^{2}}.
\label{h13}
\end{equation}
Then, the integral equation (\ref{IntEq}) becomes
\begin{equation}
\int R(\vec{n}_{1},\vec{n}_{3})\psi _{i}(\vec{n}_{1})d\vec{n}_{1}=\lambda
_{i}\psi _{i}(\vec{n}_{3})  \label{IntEq2}
\end{equation}%
with the kernel depending on the basal spins only:%
\begin{equation}
R(\vec{n}_{1},\vec{n}_{3})=\frac{\sinh \left( \beta h_{a}\right) }{\beta
h_{a}}\exp \left[ -\beta \alpha \vec{n}_{1}\cdot \vec{n}_{3}+\frac{1}{2}%
\beta \vec{h}\cdot \left( \vec{n}_{1}+\vec{n}_{3}\right) \right].
\label{R}
\end{equation}
\eqref{IntEq2} implies that the calculation of the
thermodynamics of the delta-chain is reduced to the thermodynamics
of the basal spin chain with special form of interactions, which
depend on the temperature.

Now we choose the coordinate system so that the magnetic field is
directed along the $Z$ axis. Then, $\vec{h}=\left( 0,0,h\right) $,
and unit vectors $\vec{n}$ have components $\left( \sin \theta \cos
\varphi ,\sin \theta \sin \varphi ,\cos \theta \right)$, thus
\begin{equation}
\vec{n}_{1}\cdot \vec{n}_{3}=\cos \theta _{1}\cos \theta _{3}+\sin \theta
_{1}\sin \theta _{3}\cos \left( \varphi _{1}-\varphi _{3}\right)
\end{equation}%
and the effective magnetic field (\ref{h13}) is
\begin{equation}
h_{a}=\sqrt{2+h^{2}+2\vec{n}_{1}\cdot \vec{n}_{3}+2h\left( \cos \theta
_{1}+\cos \theta _{3}\right) } \ . \label{h13sqr}
\end{equation}
Now we notice that the kernel $R$ in \eqref{R} contains the
azimuthal angles $\varphi _{1}$, $\varphi _{3}$ only as a
difference $\left( \varphi _{1}-\varphi _{3}\right) $. Then we
substitute for the eigenfunctions:
\begin{equation}
\psi _{i}\left( \vec{n}_{j}\right) =e^{im\varphi _{j}}\phi _{m,i}\left(
\theta _{j}\right)
\end{equation}%
and in terms of $x_{j}=\cos \theta _{j}$ the integral equation
(\ref{IntEq2}) becomes
\begin{equation}
\int_{-1}^{1}K_{m}\left( x_{1},x_{3}\right) \phi _{m,i}\left( x_{1}\right)
dx_{1}=\lambda _{m,i}\phi _{m,i}\left( x_{3}\right)  \label{IntEq3}
\end{equation}%
with symmetric kernel defined by an integral over $\varphi =\left(
\varphi _{1}-\varphi _{3}\right) $:
\begin{equation}
K_{m}\left( x_{1},x_{3}\right) =\int_{0}^{2\pi }\frac{d\varphi }{4\pi }%
e^{im\varphi }R_{13}\left( \varphi ,x_{1},x_{3}\right).
\label{Km}
\end{equation}
The largest eigenvalue is always given by $m=0$. The states with
$m>0$ become relevant in calculations of transverse correlation
functions \cite{Blume}, which we do not consider here. Therefore,
below we put $m=0$.

Thus, the thermodynamics of the delta-chain in the magnetic field
(\ref{H}) is reduced to the integral equation (\ref{IntEq3}) over
one variable, which can easily be calculated numerically.
Numerical results of \eqref{IntEq3} will be discussed in the
next sections.

\subsection{Classical $\Delta$-chain in a magnetic field
at low temperature}

In general, \eqref{IntEq3} completely describes the
thermodynamics of spin delta-chain in a magnetic field (\ref{H}).
However, in this Section we focus on the low temperature limit,
where explicit analytical results for the magnetization curve are
possible.

At $t\to 0$ the integration in \eqref{IntEq2} can be carried
out using the saddle point method. For this aim we need to expand
the kernel $R$ in \eqref{R} near its maximum. At first we
notice that the effective magnetic field on the apical spin in the
ground state is:
\begin{eqnarray}
h_{gs} &=&2+h,\qquad \alpha \leq \frac{1}{2}  \nonumber \\
h_{gs} &=&\frac{1}{\alpha }+\frac{1+2\alpha }{2\alpha }h,\qquad
\alpha > \frac{1}{2}.  \label{h13gs}
\end{eqnarray}
As follows from \eqref{h13gs}, $h_{gs}$ is of order of unity,
except the case $\alpha \to \infty $ and $h=0$, which is not
considered here. Therefore, in the low-temperature limit $\beta
h_{a}\gg 1$ and one can neglect the second term in $\sinh \left(
\beta h_{a}\right) $. Similarly, the denominator in \eqref{R}
can be substituted by its ground state value, so that the kernel
$R$ in the saddle point approach is approximated as
\begin{equation}
R\approx \frac{\exp \left( -\beta \mathcal{H}_{13}\right) }{2\beta
h_{gs}}, \label{RlowT}
\end{equation}%
where%
\begin{equation}
\mathcal{H}_{13}=-h_{a}+\alpha \vec{n}_{1}\cdot
\vec{n}_{3}-\frac{1}{2}\vec{h}\cdot \left(
\vec{n}_{1}+\vec{n}_{3}\right).  \label{H13}
\end{equation}
This implies that in the low-$t$ limit the behavior of the delta
chain system is described by the special form of the Hamiltonian
acting on the basal chain only:
\begin{equation}
\mathcal{H}_{\mathrm{eff}}=-\sum \sqrt{2+2\vec{n}_{2i-1}\cdot
\vec{n}_{2i+1}+2h\left( n_{2i-1}^{z}+n_{2i+1}^{z}\right)
+h^{2}}+\alpha \sum \vec{n}_{2i-1}\cdot \vec{n}_{2i+1}-h\sum
n_{2i-1}^{z}.  \label{Heff}
\end{equation}
The integral equation (\ref{IntEq2}) with the approximate
expression for kernel (\ref{RlowT}) has the form:
\begin{equation}
\int \frac{\exp \left( -\beta \mathcal{H}_{13}\right) }{2\beta
h_{gs}}\psi \left( \vec{n}_{1}\right) d\vec{n}_{1}=\lambda \psi
\left( \vec{n}_{3}\right). \label{IntEq4}
\end{equation}
The saddle point of \eqref{IntEq4} corresponds to the ground
state of the local Hamiltonian $H_{13}$. Since the ground state of
$H_{13}$ is different in the regions $\alpha \leq \frac{1}{2}$ and
$\alpha >\frac{1}{2}$, it is necessary to study these cases
separately.

\subsubsection{Ferromagnetic region and critical point $\alpha \leq
\frac{1}{2}$}
\label{fm_region}

In the ferromagnetic region $\alpha <\frac{1}{2}$ including the
vicinity of the critical point $\alpha =\frac{1}{2}$ at low $t$
nearest neighbor spins $\vec{n}_{1}$ and $\vec{n}_{3}$ are almost
parallel. In the pure ferromagnetic case $\alpha =0$ the angle
between the neighboring spin vectors is of the order of $t^{1/2}$ and
the magnetic field scales as $h\sim t^2$ \cite{universality}. As
was pointed in Ref.~\cite{DKRS}, near the critical point the
critical properties change so that the angle between the neighboring
spin vectors is of the order of $t^{1/4}$ and as will be shown
below the magnetic field scales as $h\sim t^{3/2}$ in the low-$t$
limit. Using these facts we expand the effective magnetic field
acting on the apical spin as:
\begin{equation}
h_{a}\approx 2-\frac{1}{2}\left( 1-\vec{n}_{1}\cdot \vec{n}_{3}\right) -%
\frac{1}{16}\left( 1-\vec{n}_{1}\cdot \vec{n}_{3}\right) ^{2}+\frac{1}{2}%
\vec{h}\cdot \left( \vec{n}_{1}+\vec{n}_{3}\right).  \label{h13cr}
\end{equation}
This results in the following effective local Hamiltonian (\ref{H13})%
\begin{equation}
\mathcal{H}_{13}=\left( \frac{1}{2}-\alpha \right) \left( 1-\vec{n}_{1}\cdot \vec{n}%
_{3}\right) +\frac{1}{16}\left( 1-\vec{n}_{1}\cdot \vec{n}_{3}\right) ^{2}-%
\vec{h}\cdot \left( \vec{n}_{1}+\vec{n}_{3}\right).  \label{H13cr}
\end{equation}
Though the second term in \eqref{H13cr} is of
second order in the small parameter $\left( 1-\vec{n}_{1}\cdot
\vec{n}_{3}\right) $, it becomes relevant in the vicinity of
the critical point when the factor $\left( \frac{1}{2}-\alpha
\right) $ at the first order term is small.

Next, we can simplify \eqref{IntEq4} by substituting $h_{gs}=2$
from \eqref{h13gs}, and expanding the exponent with the
magnetic field term ($\beta h\ll 1$):
\begin{equation}
\exp \left( \beta h\left( n_{1}^{z}+n_{3}^{z}\right) \right) \approx 1+\beta
h\left( n_{1}^{z}+n_{3}^{z}\right) \approx 1+2\beta hn_{3}^{z}
\end{equation}%
which transforms \eqref{IntEq4} to the form%
\begin{equation}
\left( 1+2\beta hn_{3}^{z}\right) \int e^{-\beta \mathcal{H}\left(
\vec{m}\right) }\psi \left( \vec{n}_{3}+\vec{m}\right)
d\vec{m}=4\beta \lambda \psi \left( \vec{n}_{3}\right) ,
\label{IntEq5}
\end{equation}%
where
\begin{equation}
\mathcal{H}\left( \vec{m}\right) =\frac{1-2\alpha }{4}\vec{m}^{2}+\frac{1}{64}(\vec{m}%
^{2})^{2}  \label{Hm}
\end{equation}%
and
\begin{equation}
\vec{m}=\vec{n}_{1}-\vec{n}_{3}
\end{equation}%
is a small vector of length $\left\vert \vec{m}\right\vert \sim
t^{1/4}$ which can be considered as a 2D vector ($m_{1}$,$m_{2}$)
in the plane perpendicular to the spin vector $\vec{n}_{3}$.

Now we expand the function $\psi $ in \eqref{IntEq5} to the
second order in $\vec{m}$:
\begin{equation}
\psi \left( \vec{n}+\vec{m}\right) =\psi \left( \vec{n}\right) +m_{i}\frac{%
\partial \psi \left( \vec{n}\right) }{\partial n_{i}}+\frac{1}{2}m_{i}m_{j}%
\frac{\partial ^{2}\psi \left( \vec{n}\right) }{\partial
n_{i}\partial n_{j}}, \label{psi}
\end{equation}%
where derivatives are taken along two orthogonal directions in
the plane perpendicular to the spin vector $\vec{n}$.

The Hamiltonian (\ref{Hm}) is a function of $\vec{m}^{2}$.
Therefore, linear terms in $m_{i}$ and terms $\sim m_{1}m_{2}$ in
\eqref{psi} vanish after integration over $\vec{m}$ in the integral
equation (\ref{IntEq5}). As a result, the integral equation
(\ref{IntEq5}) becomes
\begin{equation}
\left( 1+2\beta hn^{z}\right) \psi \left( \vec{n}\right) \int e^{-\beta
\mathcal{H}\left( \vec{m}\right) }d\vec{m}+\frac{1}{4}\frac{\partial ^{2}\psi }{%
\partial n_{i}^{2}}\int e^{-\beta \mathcal{H}\left( \vec{m}\right) }\vec{m}^{2}d\vec{m}%
=4\beta \lambda \psi \left( \vec{n}\right),  \label{IntEq6}
\end{equation}%
where we omit the next-order terms $\sim \beta h\vec{m}^{2}$. Now
we notice that
\begin{equation}
\frac{\partial ^{2}}{\partial n_{1}^{2}}+\frac{\partial ^{2}}{\partial
n_{2}^{2}}=-\hat{L}^{2}
\end{equation}%
is nothing but the angular momentum operator. Therefore, we come
to the Schr\"{o}dinger equation for the quantum rotator in the
gravitational field
\begin{equation}
\left( \frac{1}{2}\hat{L}^{2}-gn_{z}\right) \psi \left(
\vec{n}\right) =\mu \psi \left( \vec{n}\right),  \label{Sch1}
\end{equation}%
where the gravitational field%
\begin{equation}
g=\frac{A}{B}\beta h  \label{g}
\end{equation}%
depends on the Hamiltonian $\mathcal{H}\left( \vec{m}\right) $ via
the integrals $A$ and $B$:
\begin{eqnarray}
A &=&\int e^{-\beta \mathcal{H}\left( \vec{m}\right) }d\vec{m}  \nonumber \\
B &=&\frac{1}{4}\int e^{-\beta \mathcal{H}\left( \vec{m}\right)
}\vec{m}^{2}d\vec{m} \label{AB}
\end{eqnarray}%
and the partition function $\lambda $ is given by the lowest
eigenvalue $\mu_0$ by the equation
\begin{equation}
\lambda=\frac{A-2B\mu_0}{4\beta} .  \label{mu}
\end{equation}

The normalized magnetization is given by the scaling function
$M(t,h)=\phi(g)$, where $\phi(g)$ is determined from the ground
state energy $\mu_0$ of \eqref{Sch1} by the relation
\cite{universality}
\begin{equation}
\phi(g)=-\frac{d\mu_0}{dg} . \label{phig}
\end{equation}
The expansion of the function $\phi(g)$ for small and large $g$ as
well as the numerical calculation of $\phi(g)$ was obtained
in Ref.~\cite{universality}. It was shown in Ref.~\cite{Marko} that
the function $\phi(g)$ is well described by the approximate
equation
\begin{equation}
g=\phi(g)-\frac{1}{4}+\frac{1}{4\left( 1-\phi(g)\right) ^{2}} .
\label{Mg}
\end{equation}
Calculating $A$ and $B$ in \eqref{AB} for $\mathcal{H}\left(
\vec{m}\right) $ given by \eqref{Hm}) we have
\begin{equation}
g=\frac{h}{2t^{3/2}}f(y)  , \label{g2}
\end{equation}%
where
\begin{equation}
f(y)=\left[ \frac{e^{-y^{2}}}{\sqrt{\pi }[1+\text{erf}(y)]}+y\right] ^{-1}
\label{fy}
\end{equation}%
is the scaling function of the scaling parameter%
\begin{equation}
y=\frac{2\alpha -1}{\sqrt{t}},  \label{y}.
\end{equation}

\begin{figure}[tbp]
\includegraphics[width=4in,angle=0]{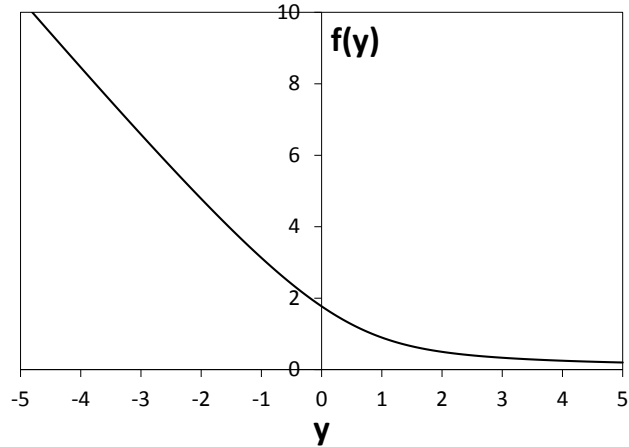}
\caption{Scaling function $f(y)$ given by \eqref{fy}.}
\label{Fig_fy}
\end{figure}

\noindent \eqref{fy} represents the analytical expression for the scaling
function $f(y)$ shown in Fig.~\ref{Fig_fy}. This function defines
the magnetization curve and the zero-field susceptibility
\begin{equation}
\chi (t,\alpha )=\frac{1}{3t^{3/2}}f(y) . \label{chi_cl}
\end{equation}
The behavior of the scaling function $f(y)$ defines two regions
with different types of thermodynamics. The first region
corresponds to the limit $y\to -\infty$, where the scaling
function $f(y)$ tends to the asymptotic $f(y)=-2y$ and the
gravitational field $g$ is
\begin{equation}
g=\frac{1-2\alpha }{t^2}h . \label{gclF}
\end{equation}
This region is limited by the condition $(1-2\alpha)\gg \sqrt{t}$
and extends up to the pure ferromagnetic case $\alpha=0$.
Therefore, we name this region as `ferromagnetic' regime. The
thermodynamics in the `ferromagnetic' regime is similar to that
for the ferromagnetic chain. In particular, the zero-field
susceptibility behaves as $\chi \sim t^{-2}$.

The second region is located near the critical point $\alpha
=\frac{1}{2}$ and is restricted by the condition $|1-2\alpha|\ll
\sqrt{t}$ ($|y|\ll 1$). In this `critical point' region one can
take the limit $f(0)=\sqrt{\pi}$ and the gravitational field
becomes
\begin{equation}
g=\frac{\sqrt{\pi}}{2t^{3/2}}h .\label{gcrF}
\end{equation}
The thermodynamics in this region is governed by the critical
point. In particular, the zero-field susceptibility behaves as
$\chi \sim t^{-3/2}$. The crossover between these two regimes
takes place at the value $y\simeq -1$, or $t\simeq
t_0=(1-2\alpha)^2$.

If we study the low-$t$ thermodynamics of the classical
$\Delta$-chain for some fixed value of $\alpha$ (not far from the
transition point), the above two regimes will manifest as follows.
The `ferromagnetic' regime taking place at very low temperatures
$t\ll t_0$ will gradually be replaced by the `critical point'
regime for $t\gg t_0$ (but still $t\ll 1$).

\begin{figure}[tbp]
\includegraphics[width=4in,angle=0]{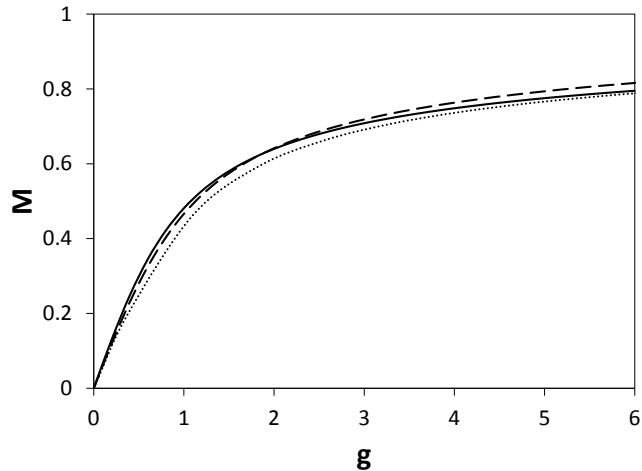}
\caption{Magnetization curve obtained by numerical solution of
the integral equation (\ref{IntEq3}) and plotted as a function of the
scaled magnetic gield $g$ (\ref{g2}) for $\alpha =0.2$ (dotted
line) and $\alpha =0.5$ (dashed line) and $t=0.1$ in comparison
with the scaling function $\phi(g)$ (solid line) representing the
exact result in the $t\to 0$ limit.} \label{Fig_Mg}
\end{figure}

The scaling function $\phi(g)$ describes the magnetization in
$t\to 0$ limits. However, the comparison of the exact numerical
solution of \eqref{IntEq3} for $\alpha =0.2$ and $\alpha =0.5$
with the scaling function given by \eqref{Mg} shows a good
agreement of both results even for $t=0.1$ as shown in
Fig.~\ref{Fig_Mg}.

\begin{figure}[tbp]
\includegraphics[width=4in,angle=0]{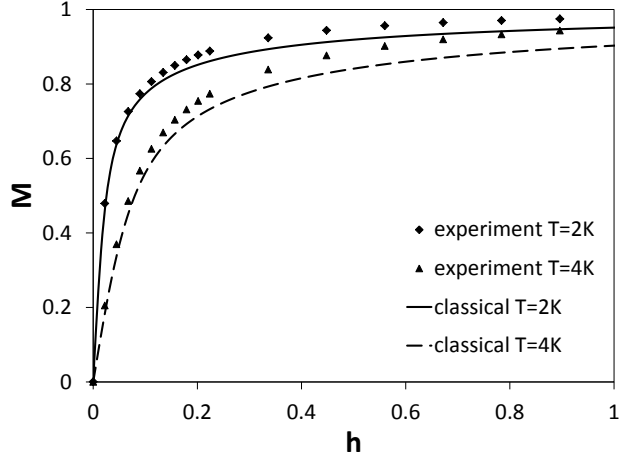}
\caption{Magnetization curve obtained by numerical solution of the
integral equation (\ref{IntEq3}) for $\alpha =0.45$ in comparison
with the experimental data for \FeGd\ for $T=2$~K and
$T=4$~K. The real magnetic field $B$ is converted to the normalized
one $h$ by equation $h=g\mu_{B}B/Jsk_B$ and the normalized
temperature relates to real temperature by $t=T/Js(s+1)$ with
$s=3$ and $J=2$~K \cite{S60}.} \label{Fig_Gd10Fe10}
\end{figure}

The comparison of the classical with the experimental
magnetization curves for \FeGd\ is shown in
Fig.~\ref{Fig_Gd10Fe10}. We find a reasonable agreement. The
slight differences between the theoretical and the experimental
curves can be attributed to quantum effects and to different
apical and basal spins present in \FeGd.

\subsubsection{Ferrimagnetic region $\alpha>\frac{1}{2}$}

In the ferrimagnetic region the neighboring basal spin vectors
form an angle $2\theta_0$ in the ground state as shown in
Fig.~\ref{Fig_ferri}. In the vicinity of the transition point
$(\alpha-\frac{1}{2})\ll 1$, the angle $\theta_0\ll 1$
(\eqref{theta0}), so that the ground state is close to the
ferromagnetic one. In this case the approach developed in the
previous subsection remains valid. This means that on the
ferrimagnetic side of the transition point (and close to it) the
magnetization curve is given by the same scaling function
$\phi(g)$ with $g$ defined by \eqref{g2}. The behavior of the
scaling function $f(y)$ for $y>0$ exhibits two low-$t$ regimes. The
`critical point' regime discussed in the previous subsection
extends to the ferrimagnetic region and is restricted by the
condition $(2\alpha-1)\ll \sqrt{t}$ ($y\ll 1$). In the limit $y\gg
1$ the scaling function behaves as $f(y)\sim 1/y$, which means
that for very low temperature $t\ll(2\alpha-1)^2$ the system is in
the `ferrimagnetic' regime with different thermodynamic exponents.
In particular, the temperature dependence of the susceptibility in
this case is $\chi\sim t^{-1}$.

\begin{figure}[tbp]
\includegraphics[width=4in,angle=0]{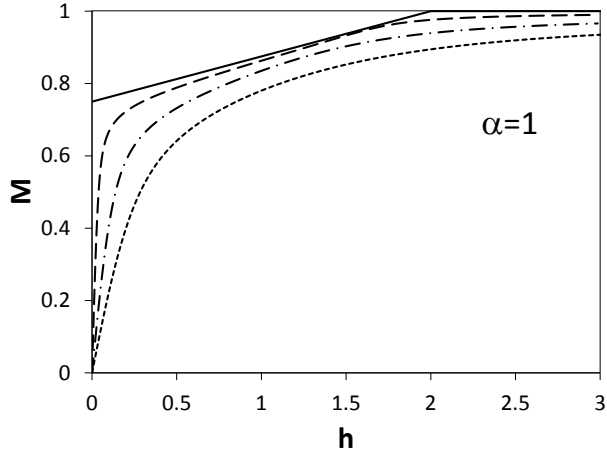}
\caption{Magnetization curves for $\alpha =1$ and several
temperatures $t=0.03$ (dashed line), $t=0.1$ (dotted-dashed line),
$t=0.2$ (dotted line), obtained by numerical solution of the integral
equation (\ref{IntEq3}). The ground state magnetization curve
(\ref{Mgs}) is shown by solid line. } \label{Fig_Mh_a1}
\end{figure}

We stress that the above scaling approach is valid in the vicinity
of the transition point only, where $\theta_0\ll 1$. Far from the
transition point the angle $\theta_0$ is no longer small, and in
order to describe the low temperature thermodynamics one needs to
expand the local Hamiltonian near the ferrimagnetic ground state
configuration described by \eqref{theta0}. The magnetization
curve in the ferrimagnetic ground state (\ref{Mgs}) and for several small
values of $t$ for $\alpha=1$
is shown in Fig.~\ref{Fig_Mh_a1}. As can be seen the magnetization curves
approach the ground state curve with decreasing $t$.
According to Fig.~\ref{Fig_Mh_a1} the magnetization curves have
three different scales in the magnetic field which should be
studied separately: $h\ll t$, $t<h<h_{\mathrm{sat}} $, and $h\geq
h_{\mathrm{sat}}$.

For very low magnetic field, $h\ll t$, the ground state spin
configurations can be described in terms of finite step random
walk on the unit sphere in a weak gravitational field \cite{DKRS}.
The magnetization in this case increases linearly with the
magnetic field and the zero field susceptibility was
calculated in Ref.~\cite{DKRS}:%
\begin{eqnarray}
M &=&\chi h  \label{M_chi_h} \\
\chi &=&\frac{1}{6t}\frac{2\alpha +1}{2\alpha -1}.
\end{eqnarray}
Then, for higher magnetic field $t<h<h_{\mathrm{sat}}$ the
magnetization approaches its ground state value (\ref{Mgs}), and
the integral equation (\ref{IntEq4}) can be solved using the saddle
point approximation. For this aim we introduce small deviations
$\delta_1,\delta_3,\varepsilon $ from the ferrimagnetic
ground state (\ref{theta0}):%
\begin{eqnarray}
\cos \theta _{1} &=&\cos \theta _{0}+\delta _{1}  \nonumber \\
\cos \theta _{3} &=&\cos \theta _{0}+\delta _{3}  \nonumber \\
\varphi &=&\pi +\varepsilon.
\end{eqnarray}
The leading terms of the expansion of the local Hamiltonian in
$\delta_1,\delta_3,\varepsilon $ is:
\begin{equation}
\mathcal{H}_{13}=E_{gs1}+A_{1}(\delta _{1}^{2}+\delta
_{3}^{2})+2B_{1}\delta _{1}\delta _{3}+C_{1}\varepsilon ^{2},
\end{equation}%
where%
\begin{eqnarray}
E_{gs1} &=&-\frac{(2+h)^{2}}{8\alpha }-h-\alpha \\
A_{1} &=&\frac{8\alpha ^{3}(2\alpha h+h+2)-\alpha (h+2)^{2}}{[16\alpha
^{2}-(h+2)^{2}](2\alpha h+h+2)} \\
B_{1} &=&A_{1}-\frac{\alpha h(2\alpha +1)(h+2)^{2}}{[16\alpha
^{2}-(h+2)^{2}](2\alpha h+h+2)} \\
C_{1} &=&\frac{h(2\alpha +1)[16\alpha ^{2}-(h+2)^{2}]}{32\alpha (2\alpha
h+h+2)}
\end{eqnarray}
The solution of the integral equation (\ref{IntEq4}) in this case is%
\begin{equation}
\lambda =\exp \left( -\frac{E_{gs1}}{t}\right) \frac{\pi t^2}{h_{gs}%
\sqrt{A_{1}C_{1}+C_{1}\sqrt{A_{1}^{2}-B_{1}^{2}}}}.
\label{lambda1}
\end{equation}
The magnetization is given by the relation:%
\begin{equation}
M=t\frac{\partial \ln \lambda}{\partial h}. \label{Mth}
\end{equation}%
The magnetization curve approaches the ground state expression (\ref{Mgs})
in low-$t$ limit by the law:%
\begin{equation}
M=M_{\mathrm{gs}}-t f_M\left( h,\alpha \right),
\end{equation}%
where the explicit form of the function $f_M\left( h,\alpha
\right)$ is very cumbersome and we do not present it here.

\begin{figure}[tbp]
\includegraphics[width=4in,angle=0]{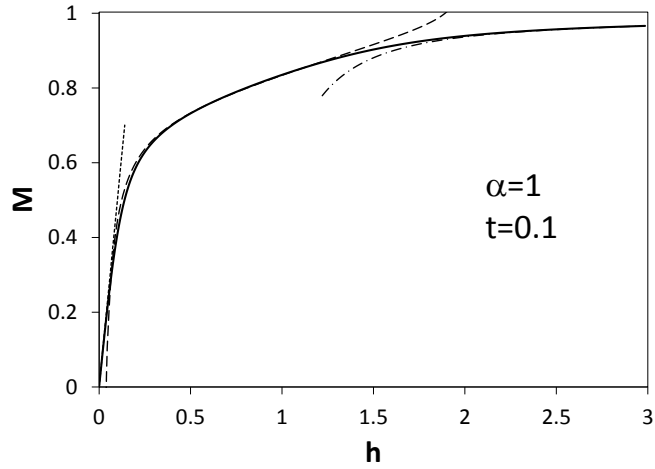}
\caption{Magnetization curves for $\alpha =1$ and $t=0.1$ obtained
by numerical solution of the integral equation (\ref{IntEq3}) (solid
line) and approximate equations (\ref{M_chi_h}), (\ref{lambda1})
and (\ref{lambda2}) in the corresponding regions (dashed lines).}
\label{Fig_Mh_a1_appr}
\end{figure}

Finally, when the magnetic field is higher than the saturation
one, $h>h_{\mathrm{sat}}$, the ground state becomes ferromagnetic
and the magnetization only slightly differs from its fully
saturated value. That means that the angles $\theta_1$ and
$\theta_3$ are small and the expansion of the local Hamiltonian
becomes:
\begin{eqnarray}
\mathcal{H}_{13} &=&E_{gs2}+A_{2}(\theta _{1}^{2}+\theta
_{3}^{2})+2B_{2}\theta
_{1}\theta _{3}\cos \varphi \\
E_{gs2} &=&-2+\alpha -2h \\
B_{2} &=&\frac{1}{2}\left( \alpha -\frac{1}{2+h}\right) \\
A_{2} &=&B_{2}+\frac{h-h_{\mathrm{sat}}}{4}.
\end{eqnarray}
In this case after some algebra the solution of the integral equation
(\ref{IntEq4}) yields the partition function%
\begin{equation}
\lambda =\frac{1}{8A_{2}}\frac{t^2}{2+h}\exp \left( -\frac{E_{gs2}}{t}%
\right) \left( 1+\frac{B_{2}^{2}}{4A_{2}^{2}}\right)  . \label{lambda2}
\end{equation}
As shown in Fig.~\ref{Fig_Mh_a1_appr} for $\alpha =1$
Eqs.~(\ref{M_chi_h}), (\ref{lambda1}), (\ref{Mth}) and
(\ref{lambda2}) perfectly describe the magnetization curve in the
corresponding regions of the magnetic field.

\section{Quantum effects}

In the preceding Section we represented results for the classical
delta-chain in the magnetic field. Since the classical model
corresponds to the limit $s\to\infty $, a natural question arises
about the relation of the classical results to those of the
quantum spin-$s$ model (\ref{Hq}). In this respect it is important
to mention Ref.~\cite{universality} where it was conjectured that
the magnetization curves of the quantum and classical
ferromagnetic chain coincide in the low-temperature limit and
described by an universal function $\phi (g_{F})$
(\eqref{phig}) of the scaling variable $g_{F}$
\begin{equation}
g_{F}=\frac{s^3 H}{T^2}.\label{gsF}
\end{equation}
In this Section we will use non-renormalized temperature
$T=ts^{2}$ and the magnetic field $H=sh$. As the ferromagnetic
chain corresponds to the particular case $\alpha =0$ of our model,
the problem of `universality' of the classical results for $\alpha
>0$ will be in the focus of our attention. Additional motivation
to study the quantum effects to the classical results is that
\FeGd\ is described by the quantum model with
relatively high but nevertheless finite spin values. For the
analysis of the magnetic properties of the quantum spin model we
investigate finite chains imposing periodic boundary conditions
using the numerical exact diagonalization (ED) \cite{spinpack} and
the finite-temperature Lanczos (FTL) technique \cite{FTL1,FTL2}.

\subsection{Transition point}

We start our analysis from the transition point $\alpha
=\frac{1}{2}$. The spin-$\frac{1}{2}$ case of quantum model
(\ref{Hq}) at the transition point was studied in detail in
Ref.~\cite{KDNDR}. It was shown that this model has many very
specific properties: a flat one-magnon spectrum, localized
one-magnon states and multi-magnon complexes, a macroscopic
degeneracy of the ground state and a residual entropy,
exponentially low-lying excitations, a multi-scale structure of
the energy spectrum \cite{DK15}. It turns out that all these
specific properties of the spin-$\frac{1}{2}$ model carry over to
the models with higher values of spin with some inessential
modifications, which we will briefly describe below.

The ground state of the quantum delta-chain with any value of $s$
at the critical point $\alpha =\frac{1}{2}$ consists of exact
multi-magnon bound states exactly like the $s=\frac{1}{2}$ model
and the number of the ground states, $B_{N/2}^{k}$, for fixed
value $S^{z}=S_{\max }-k$ ($S_{\max }=sN$) is \cite{KDNDR}
\begin{eqnarray*}
B_{N/2}^{k} &=&C_{N/2}^{k},\qquad 0\leq k\leq \frac{N}{4},\qquad 2S_{\max }-%
\frac{N}{4}<k\leq 2S_{\max } \\
B_{N/2}^{k} &=&C_{N/2}^{N/4},\qquad \frac{N}{4}+1\leq k\leq
2S_{\max }-\frac{N}{4}
\end{eqnarray*}%
where $C_{m}^{n}=\frac{m!}{n!(m-n)!} $ is the binomial
coefficient.

The contribution to the partition function from only these
degenerate ground states is
\begin{equation}
Z_{GS}=\sum_{k}B_{N/2}^{k}\exp (\frac{(S_{\max }-k)H}{T}) .
\label{Zgs}
\end{equation}
Using a saddle-point approximation to estimate of \eqref{Zgs} we
obtain the corresponding normalized magnetization in the form
\begin{equation}
M_{GS}=1-\frac{1}{2s[1+\exp (H/T)]}  \label{mgs}
\end{equation}
As follows from \eqref{mgs}, the magnetization at the critical
point for $H \to 0$ is
\begin{equation}
M_{GS}=1-\frac{1}{4s}  \label{m_ds}
\end{equation}%
and it changes from $M_{GS}=\frac{1}{2}$ for $s=\frac{1}{2}$ to
$M_{GS}=1$ for the classical limit $s\to\infty $.

\begin{figure}[tbp]
\includegraphics[width=4in,angle=0]{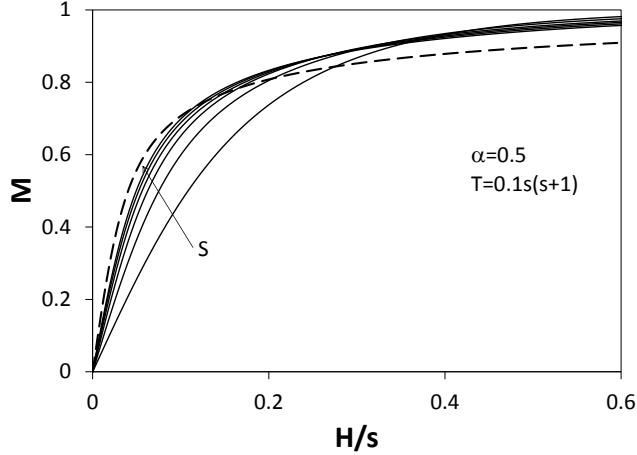}
\caption{Magnetization curves $M(H/s)$ for the quantum models with
$s=\frac{1}{2}$ ($N=36$), $s=1,\frac{3}{2}$ ($N=16$),
$s=2,\frac{5}{2},3$ ($N=12$) (solid lines) at the transition point
$\alpha =\frac{1}{2}$ for $T/s(s+1)=0.1$. The magnetization curve
of the classical model for $\alpha =\frac{1}{2}$ and $t=0.1$ is
shown by the dashed line.} \label{Fig_Mh_a05}
\end{figure}

\begin{figure}[tbp]
\includegraphics[width=4in,angle=0]{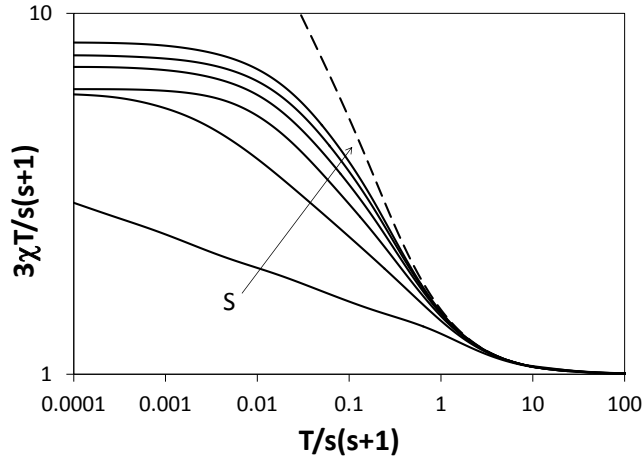}
\caption{Dependence of $3\chi T/s(s+1)$ on the normalized temperature
$T/s(s+1)$ for classical (dashed line) and quantum spin-$s$ (solid lines) delta-chain
calculated at the critical point $\alpha=0.5$.}
\label{Fig_chi_T_S}
\end{figure}

According to \eqref{m_ds} the magnetization $M_{GS}$ is finite
for $H \to 0$, which would clearly contradict the statement that long range
order cannot exist in one-dimensional systems at $T>0$. For the
correct description of $M(H,T)$ it is thus necessary to take into
account the full spectrum of the model. Unfortunately, such
analytical calculation is impossible, and we therefore carried
out ED and FTL calculations of $M(H,T)$ for different values of $s$ and
$N$. Corresponding results together with that for the classical
model are shown in Fig.~\ref{Fig_Mh_a05}. As it can be seen from
Fig.~\ref{Fig_Mh_a05} the behaviors of the classical and quantum
model are very different. It implies that there is no universality
at the critical point. At the same time, there is one interesting
point related to the behavior of the magnetization at low magnetic
field. It was shown in Ref.~\cite{KDNDR} that the magnetization of
the $s=\frac{1}{2}$ delta-chain is $M\sim H/T^{\gamma }$ with an
exponent $\gamma =1.09$. On the other hand in the classical model
($s\to\infty $) $\gamma =\frac{3}{2}$ according to
\eqref{chi_cl}. Therefore, it can be expected that the exponent
$\gamma $ is a function of $s$. To clarify this point we have
calculated the zero-field susceptibility $\chi $ for different $s$
and $N$. The dependencies $\chi (T)$ are shown in
Fig.~\ref{Fig_chi_T_S} as log-log plot of $3\chi T/s(s+1)$ vs.
$T/s(s+1)$. The solid lines denote from bottom to top:
$s=\frac{1}{2}$ ($N=36$), $s=1$ ($N=16$) and
$s=\frac{3}{2},2,\frac{5}{2},3$ with $N=12$. The classical curve
is shown by dashed line. As it can be seen in
Fig.~\ref{Fig_chi_T_S} all curves tends to $1$ in the high
temperature limit, which is in accord with high-$T$ behavior of
the susceptibility $\chi =s(s+1)/3T$. Then, for lower temperature
all curves diverge from each other and in a definite intermediate
temperature region the curves\ have linear behavior with different
slope which implies a power-law dependence
\begin{equation}
\chi =r(s)/T^{\gamma (s)}  \label{chi1}
\end{equation}

\begin{figure}[tbp]
\includegraphics[width=4in,angle=0]{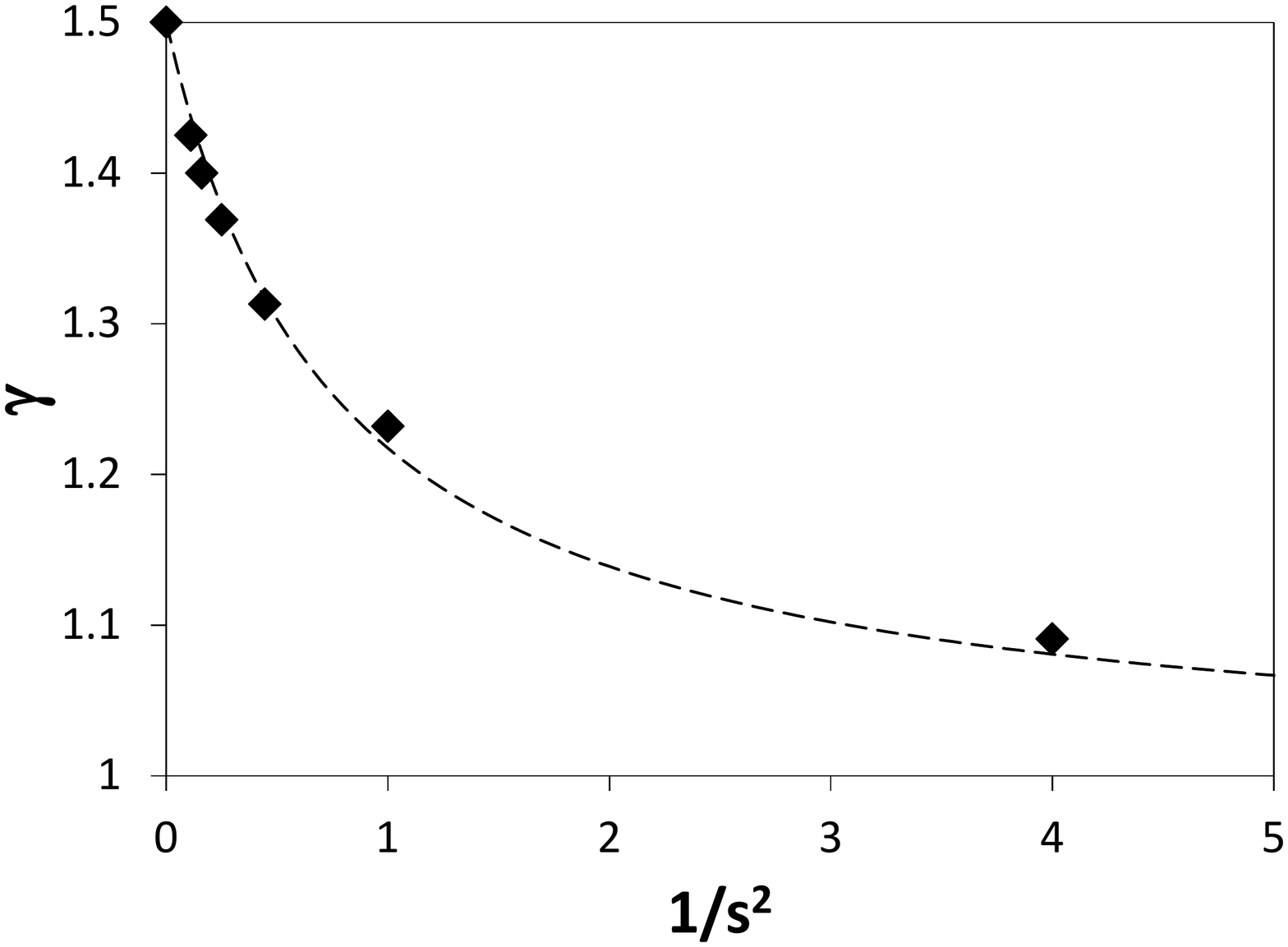}
\caption{Dependence of the critical exponent $\gamma$ on the spin value
$s$. The dashed line represents the approximate expression $\gamma
=\frac{3s^2+2.6}{2s^2+2.6}$.} \label{Fig_gammaS}
\end{figure}
That means that the low-field behavior of the magnetization is
$M\sim H/T^{\gamma (s)}$. The dependence of the critical exponent
on spin value $\gamma (s)$ is shown in Fig.~\ref{Fig_gammaS} and
it can be seen that $\gamma\to\frac{3}{2}$ in the classical limit
$s\gg 1$. As further decreasing $T/s(s+1)$ for all solid curves
the sloping part in Fig.~\ref{Fig_chi_T_S} is followed by a flat
part related to finite-size effects. At $T\to 0$ the solid curves
tend to the values determined by the contributions of the
degenerate ground states. These contributions for finite
delta-chains can be found by the calculations of the zero-field
susceptibility per spin using \eqref{Zgs}, which results in
\begin{equation}
\chi =\frac{c_{N}(s)N}{T}  \label{chi2} ,
\end{equation}%
where $c_{N}(s)=\frac{1}{2}(s-\frac{1}{4})^{2}$ for $N\gg 1$.
We suppose that both equations (\ref{chi1}) and (\ref{chi2}) for
$\chi (T)$ are described by a single finite-size scaling function
which has the form \cite{KDNDR}
\begin{equation}
\chi (T)=T^{-\gamma }F(c_{N}(s)NT^{\gamma -1})
\end{equation}

For small $x$ the function $F(x)$ gives (\ref{chi2}) and in the
thermodynamic limit $N\to\infty $ the scaling function tends to
the value $r(s)$ in accord with \eqref{chi1}. The crossover
between these two types of the susceptibility behavior occurs at
$x\simeq 1$ which defines the crossover temperature $T^{\ast }\sim
N^{-1/(\gamma -1)}$. At $T<T^{\ast }$ finite-size effects are
essential and $\chi $ is given by \eqref{chi2}. The crossover
temperature $T^{\ast }$ increases with $s$ and the region of
finite-size behavior of $\chi $ increases.

\subsection{Ferromagnetic phase}

As was noted in the beginning of this Section, in the special case
$\alpha=0$ the magnetization curves of both quantum and classical
delta-chain models coincide in the low-temperature limit.
According to the scaling hypothesis \cite{universality} the
normalized magnetization $M$ for the infinite chain is expressed
at $T\to 0$ and $\frac{H}{T}\to 0$ (but with fixed $g_F$
(\ref{gsF})) as $M(T,h)=\phi(g_F)$ and the function $\phi(g_F)$ is
obtained by calculating the eigenspectrum of the quantum rotator
Hamiltonian (\ref{Sch1}) in the gravitational field $g_F$. As
noted in Ref.~\cite{universality} the hypothesis of
universality originates in the universal behavior of the spin-wave
excitations above the ferromagnetic ground state in both quantum
and classical models. Similarly to the case $\alpha =0$ one can
expect that such universality remains in the ferromagnetic part of
the ground state phase diagram ($\alpha <\frac{1}{2}$) with $g_F$
in \eqref{gsF} being replaced by
\begin{equation}
g_F = \frac{(1-2\alpha) s^3 H}{T^2}.\label{gs}
\end{equation}
in accordance with \eqref{gclF} for the classical model.

The universality for $\alpha <\frac{1}{2}$ is partly confirmed by
the fact that the leading terms of the zero-field susceptibility
at $T\to 0$ for the classical model and that obtained in a frame
of the modified spin-wave theory \cite{MSWT} for the quantum model
coincide \cite{DKRS}. Unfortunately, modified spin-wave theory is
restricted to the zero magnetic field case and it can not confirm
the universality of the magnetization curve.

\begin{figure}[tbp]
\includegraphics[width=4in,angle=0]{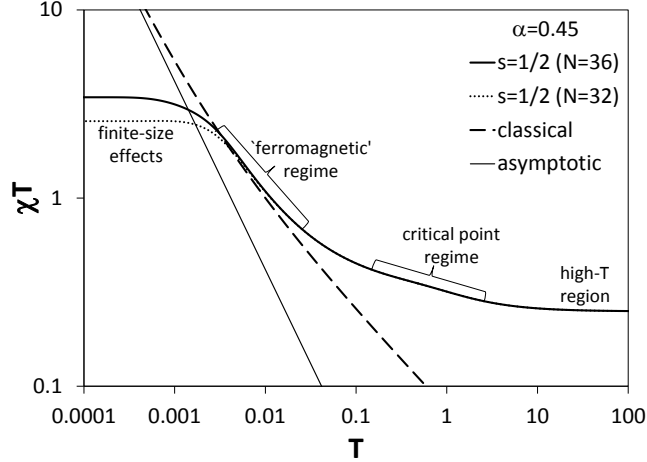}
\caption{Susceptibility times temperature, $\chi T$, in dependence
on $T$ for $\alpha=0.45$ obtained numerically by FTL for $s=1/2$
and $N=32$ (dotted line), $N=36$ (thick solid line). The classical
curve is shown by the dashed line. The thin solid line describes
low-$T$ asymptotic $\chi=(1-2\alpha)/24T^2$.}
\label{Fig_Chi_T_a_045_s12}
\end{figure}

However, the extension of the hypothesis of the universality for
the case $\alpha \neq 0$ and especially for $\alpha $ close to the
transition point $\alpha =\frac{1}{2}$ needs some comments. As it
was shown in the preceding Section the scaling parameter $g$ in
the classical model has two different forms given by
Eqs.~(\ref{gclF}) and (\ref{gcrF}) for $T\ll T_{0}$ and $T\gg
T_{0}$, respectively, where $T_0=(1-2\alpha)^2 s^2$ is the
temperature of the crossover. For $T\ll T_{0}$ this parameter
takes the form (\ref{gs}), while for $T\gg T_{0}$ it corresponds
to that for the transition point regime, where  the behavior of
the classical and quantum models is very different. Therefore, one
can expect that there is identical universality of the classical
and quantum models in the low-temperature region $T\ll T_{0}$
only.

The quantum models also have different low-temperature regimes
when $\alpha $ is close to the transition point. As an example we
show in Fig.~\ref{Fig_Chi_T_a_045_s12} the dependence of the
susceptibility for the $s=\frac{1}{2}$ delta-chain and $\alpha
=0.45$ with $N=32$ and $N=36$ obtained by FTL calculations, where for
convenience we represent this dependence as log-log plot of $\chi
T(T)$. At first we note that the curves with $N=32$ and $N=36$
perfectly coincide for $T>0.003$, which means that they correctly
describe the thermodynamic limit in this region. In the high
temperature limit the curves tend to a constant, which implies the
correct asymptotic $\chi(T)=1/(4T)$. In the temperature range
$0.1\lesssim T \lesssim 3$ the slope of the curve is very close to
that obtained for $s=\frac{1}{2}$ at the transition point
\cite{KDNDR}: $\chi(T) \sim T^{-\gamma}$ with $\gamma =1.09$.
Therefore, we refer this region to the `critical point' regime.

For temperatures lower than the `critical point' region the slope
of the curves increases and after some crossover region the
quantum curves approach the classical curve shown in
Fig.~\ref{Fig_Chi_T_a_045_s12} by a dashed line. We name the region,
where the quantum curves are close to the classical one,
$0.003\lesssim T \lesssim 0.02$, the `ferromagnetic' one. Though
the slope of the curves in this region corresponds to $\gamma\sim 1.7$
instead of a `ferromagnetic' $\gamma=2$, we see that all curves
converge to the `ferromagnetic' low-$T$ asymptotic
$\chi=(1-2\alpha)/24T^2$, shown by the thin solid line in
Fig.~\ref{Fig_Chi_T_a_045_s12}. For $T<0.003$ the quantum curves
for $N=32$ and $N=36$ diverge from each other and both from the
classical curve, establishing the `finite-size effect' region with
non-thermodynamic behavior. Looking at
Fig.~\ref{Fig_Chi_T_a_045_s12} it is natural to assume that the
quantum curve corresponding to very long chains would go further
into the lower $T$ region close to the classical curve and both
asymptotically approach the thin solid line, i.e., the ferromagnetic
law $\chi=(1-2\alpha)/24T^2$. This means that for the infinite
delta-chain the ferromagnetic region exists up to $T=0$.
Unfortunately, for $\alpha \neq 0$ the quantum models can be
studied only by numerical calculations of finite delta-chains,
which due to finite-size effects restrict the low temperatures.

\begin{figure}[tbp]
\includegraphics[width=4in,angle=0]{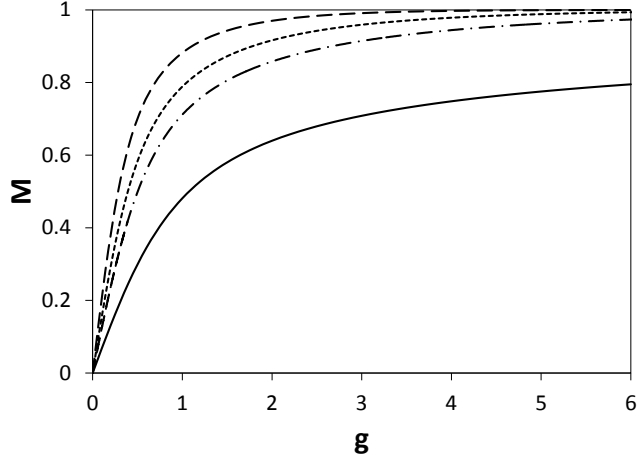}
\caption{Magnetization curve for $\alpha=0.45$ obtained
numerically by FTL for $s=1/2$, $N=36$ and three different
temperatures located in the `ferromagnetic region' of
Fig.~\ref{Fig_Chi_T_a_045_s12}: $T=0.004$ (dashed-dotted line),
$T=0.0075$ (short-dashed line), $T=0.015$ (long-dashed line). The
data are plotted as a function of the scaled magnetic field $g_F$
(\ref{gs}). The exact universal magnetization curve $\phi(g_F)$ is
shown by the solid line.} \label{Fig_Mg_a045_s12}
\end{figure}

The magnetization curves in the `ferromagnetic' temperature region
for the $s=\frac{1}{2}$ model with $N=36$ and for $\alpha =0.45$
obtained by numerical FTL calculations is shown in
Fig.~\ref{Fig_Mg_a045_s12} as a function of the
`ferromagnetically' scaled field $g_F$ (\ref{gs}). In
Fig.~\ref{Fig_Mg_a045_s12} we also show the scaling function
$\phi(g)$. As it can be seen the quantum magnetization curves tend
to the scaling function as the temperature decreases. However, the
difference between these curves and $\phi(g)$ is rather
appreciable. The point is that the function $\phi(g)$ represents
the leading term in the low-temperature expansion of the
magnetization. The temperatures corresponding to the magnetization
of the $s=\frac{1}{2}$ model in Fig.~\ref{Fig_Mg_a045_s12} are
about $T_{0}$. At such a temperature the next terms in the
low-temperature expansion of the magnetization are of the same
order as the leading term. This appreciable difference of the
initial slope of the quantum magnetization curve and $\phi(g)$ can
be also seen in Fig.~\ref{Fig_Chi_T_a_045_s12}: in the
`ferromagnetic' region the values $\chi(T)$ for quantum curve is
approximately two times larger than that for the asymptotic line
corresponding to the initial slope of $\phi(g)$. The comparison of
the classical and asymptotic lines in
Fig.~\ref{Fig_Chi_T_a_045_s12} shows that the difference would
become $\sim 10\%$ for $T\lesssim 0.0005$, but in order to avoid
the finite-size effects at such low temperatures one needs to
calculate very long chains.

In the `finite-size' region the correlation length
$\xi=(1-2\alpha)s^2/T$ is much larger than the system size
(especially for $\alpha $ close to $\frac{1}{2}$) accessible in
exact diagonalization (ED) ($N\sim 24$) or FTL ($N\sim 36$)
calculations. In this region the finite-size effects are essential
and the scaling function for the magnetization depends on two
parameters $\phi(g_{F},q)$ \cite{universality} with
\begin{equation}
q=\frac{(1-2\alpha )s^2}{TN}.
\end{equation}
At $T\to 0$ and $q\gg 1$ the function $\phi(g,q)$ is given by the
Langevin equation
\begin{equation}
M=\phi(g_{F},q)=\coth (x)-\frac{1}{x}  \label{M_L}
\end{equation}%
with%
\begin{equation}
x=\frac{g_{F}}{q}=\frac{NsH}{T}.
\end{equation}
The magnetization calculated for the quantum delta-chain at
$\alpha =0.45$ with $s=\frac{1}{2}$ and $N=36$ well agrees with
\eqref{M_L}.

The numerical calculations of the magnetization of the quantum
$s=\frac{1}{2}$ model for temperatures $T\gg T_{0}$ show
significant difference from the classical scaling function
$\phi(g)$. Therefore, we conclude that the magnetization for
$0\leq \alpha <\frac{1}{2}$ is a universal function for both
quantum and classical delta-chain only in the `ferromagnetic'
regime ($T\ll T_{0}$).

\begin{figure}[tbp]
\includegraphics[width=4in,angle=0]{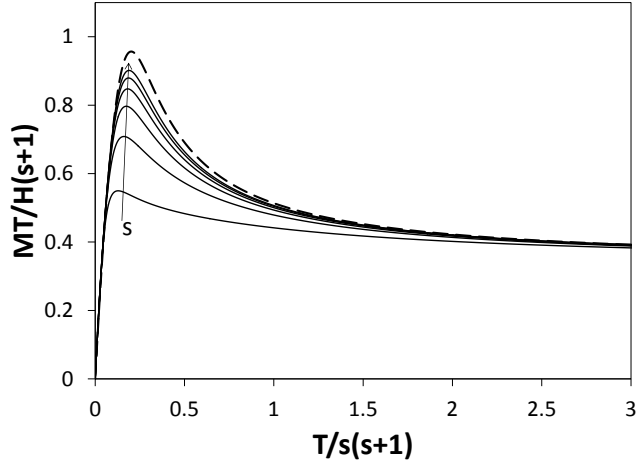}
\caption{Comparison of quantum (solid lines) and classical (dashed
line) dependencies $MT/H(s+1)$ vs. $T/s(s+1)$ calculated for
$\alpha=0.45$ and $h=0.1$.} \label{Fig_MTH}
\end{figure}

As discussed in Secs.~\ref{intro} and \ref{fm_region} the
classical approximation for F-AF delta-chain is justified
for \FeGd, because the spin quantum numbers for
Fe and Gd ions are rather large. The characteristic feature
related to the susceptibility of \FeGd\ is
a maximum in the temperature dependence of the quantity $MT/H$ in
a fixed magnetic field. The calculation of this quantity for
classical model shows good agreement with the experimental data.
In particular, the maximum $(MT/H)_{max}\sim 720$~cm$^3$K/mol is
reached at $T_{max}\sim 4$~K in comparison with experimental data
$(MT/H)_{max}\sim 745$~cm$^3$K/mol is reached at $T_{max}\sim 3$~K.
The temperature dependence of $MT/H$ for quantum models with
different values of spin $s$ is shown in Fig.~\ref{Fig_MTH}
together with that for the classical model. As it can be seen in
Fig.~\ref{Fig_MTH} the dependencies $MT/H$ approach to the
classical curve as $s$ increases.

\subsection{Ferrimagnetic phase}

\begin{figure}[tbp]
\includegraphics[width=4in,angle=0]{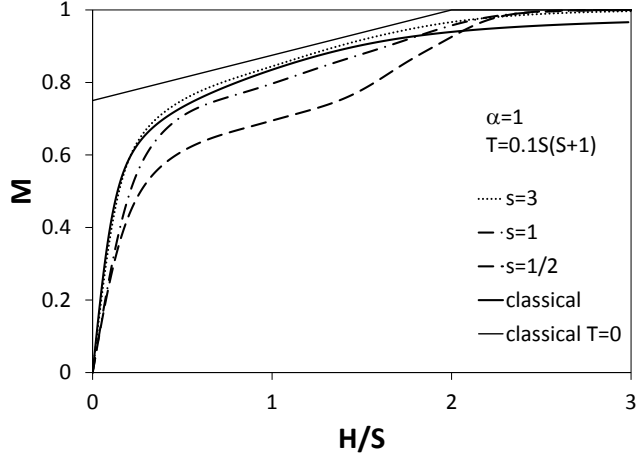}
\caption{Comparison of the magnetization curves $M(H/s)$ for the
classical (thick solid line) and quantum models with $s=1/2, N=36$
(dashed line), $s=1, N=16$ (dashed-dotted line), $s=3, N=12$
(dotted line) in the ferrimagnetic region $\alpha =1$ for
$T/s(s+1)=0.1$. The ground state magnetization curve of the
classical model (\ref{Mgs}) is shown by thin solid line.}
\label{Fig_Mhq_a1}
\end{figure}

The ground state of the classical model is ferrimagnetic at
$\alpha >\frac{1}{2}$. As we noted before, in Ref.~\cite{Tonegawa}
it was stated that a ferrimagnetic ground-state phase is also
realized for the $s=\frac{1}{2}$ quantum delta-chain. At the same
time the behavior of the magnetization curve of the classical and
quantum models is very different as it is shown in
Fig.~\ref{Fig_Mhq_a1} for $\alpha =1$. It is possible to state
with certainty that there is no universality in this phase. At
present it is not much known about the ground state phase of the
quantum models with $s>\frac{1}{2}$ and this problem needs further
study. One interesting point is the dependence of the
magnetization behavior on $s$. As it is shown in
Fig.~\ref{Fig_Mhq_a1} the magnetization curves rapidly approach to
the classical one when $s$ increases. It can be expected that the
magnetization of the quantum model in $s\gg 1$ limit will coincide
with the classical curve.

\section{Summary}

In this paper we have studied the delta-chain with competing
ferro- and antiferromagnetic interactions $J_1$ and $J_2$  in the
external magnetic field. At $\alpha=J_2/|J_1|=1/2$, this model
belongs to the class of flat-band models exhibiting a massively
degenerated ground state leading to a residual entropy. Since, a
magnetic field partially lifts the degeneracy, the influence of
the field on the low-temperature physics is tremendous.
Interestingly, there is a finite-size realization of the model,
namely the magnetic molecule \FeGd, that has $J_1$ and
$J_2$ close to the flat-band point. In the present study,  for the
classical model exact results for the thermodynamics are obtained.
It is shown that the calculation of the magnetization for $\alpha
\leq \frac{1}{2}$ in the limit $T\to 0$ and $\frac{H}{T}\to 0$
reduces to the solution of the Schr\"{o}dinger equation for the
quantum rotator in the gravitational field $g$ which depends on
the temperature. The low-temperature region of the classical model
consists of two regions $T\ll T_{0}$ and $T\gg T_{0}$ ($T_{0}\sim
(1-2\alpha)^2s^2$) with different type of the $g(T)$ dependence.
The magnetization for $T\ll T_{0}$ is a universal function of the
scaling parameter $g$ which is valid for both classical and the
quantum models. In particular, the susceptibility behaves as $\chi
\sim T^{-2}$. For $T\gg T_{0}$ the behavior of the magnetization
and the susceptibility is the same as in the critical point
$\alpha =\frac{1}{2}$ and it is different for the classical and
the quantum models. In this case the susceptibility of the
classical model behaves as $\chi \sim T^{-3/2}$ while
$\chi \sim T^{-\gamma}$ with $\gamma =1.09$ for the quantum $s=\frac{1}{2}$
model. Generally, the value of the exponent $\gamma $ depends on
$s$ and it tends to the classical value $\gamma =\frac{3}{2}$ when
$s$ increases.

We compare the obtained results with the experimental data for
\FeGd, which is a finite-size realization of the considered
model with $\alpha \simeq $0.45. We show that the
magnetization $M(H)$ of both classical and quantum model with
$s=3$ agrees well with the experimental magnetization curves
measured at $T=2$~K and $T=4$~K. We also discuss the maximum in the
temperature dependence of the quantity $MT/H$ at fixed magnetic
field and show that it agrees very well with the experimentally
observed one.

\section*{Acknowledgment}

Computing time at the Leibniz Center in Garching is gratefully
acknowledged.

\end{document}